\documentclass[%
pra,
reprint,
twocolumn,
amsmath,
amssymb,
aps,
showpacs]{revtex4}

\usepackage{natbib}
\usepackage{graphicx}
\usepackage[tight]{subfigure}
\usepackage{bm}
\usepackage{dcolumn}
\usepackage{amssymb}
\usepackage{amsbsy}
\usepackage{times}
\usepackage{color}
\usepackage{float}
\usepackage[colorlinks,bookmarks=false,citecolor=blue,linkcolor=red,urlcolor=blue]{hyperref}

\begin{document}
\title{Strong-coupling expansion for ultracold bosons in an optical lattice at finite temperatures in the presence of superfluidity}
\author{Manjari Gupta}
\author{H. R. Krishnamurthy}
\affiliation{Center For Condensed Matter Theory, Department of Physics, Indian Institute of Science, Bangalore 560012, India}
\author{J. K. Freericks}
\affiliation{Department of Physics,Georgetown University, Washington, D.C. 20057, USA}
\date{\today}

\begin{abstract}
We develop a strong-coupling ($t \ll U$) expansion technique for calculating the density profile for bosonic atoms trapped in an optical lattice with an overall harmonic trap at finite temperature and finite on site interaction in the presence of superfluid regions. Our results match well with quantum Monte Carlo simulations at finite temperature. We also show that the superfluid order parameter never vanishes in the trap due to proximity effect. Our calculations for the scaled density in the vacuum to superfluid transition agree well with the experimental data for appropriate temperatures. We present calculations for the entropy per particle as a function of temperature which can be used to calibrate the temperature in experiments. We also discuss issues connected with the demonstration of universal quantum critical scaling in the experiments.
\end{abstract}

\maketitle


\section{Introduction}

Ultracold atoms trapped in optical lattices (which are generated by the interference of counter propagating laser beams) are of great current interest \cite{Bloch} for they can emulate model Hamiltonians of importance in describing condensed matter systems with tunable interactions. This gives an opportunity to study some unsolved lattice based models over a wide range of their parameters and to understand strong correlation phenomena. It is also useful to study quantum phase transitions and quantum critical phenomena.

For example,  ultracold bosonic atoms trapped in deep optical lattices correspond to the Bose Hubbard model \cite{Fisher,Jacks}. There is also an overall harmonic trap potential given by $V_{tr} |\vec r|^2$ which causes an effective chemical potential variation throughout the lattice resulting in the coexistence of consecutive annular regions of Mott/normal and superfluid phases \cite{Bloch2}. Recent advances in imaging techniques now permit a mapping of the density profiles of ultracold atomic systems as a function of radial distance at near single site spatial resolution \cite{Chin,Chin2,Bloch2,Bakr}. It is important to benchmark these density profiles with theoretical calculations which will also in turn validate the theoretical techniques used to calculate the density profiles.

There have also been extensive quantum Monte Carlo (QMC) simulational studies of the Bose-Hubbard model with an overall harmonic trap, at zero or finite temperature \cite{Nandini,Alet}. While QMC simulations are more accurate than most analytical approaches used so far, they can be time consuming as the system size increases. Approximate analytical calculations that do not significantly compromise on the quality of the results, are of great value as they are computationally less demanding, and can provide a better understanding of the dominating physical processes for experimental parameters of interest. While there have also been many analytical approaches to the Bose-Hubbard Model \cite{Menotti,Itay,Mattias,Batrouni,Rigol,Krishnendu,Freericks2,Jamshid}, to our knowledge, none of them do a good job of giving the density distribution throughout the trap for finite temperature and large interaction strength $U$.

In this paper, we present calculations extending the finite temperature strong-coupling expansion  formalism developed in Ref. ~\onlinecite{Freericks} to the present case of trapped bosons allowing for the possibility of superfluidity. This extension is essential, for if the strong-coupling expansion (expansion in $t/U$) is carried out for a case in a trap at low temperature, then it shows diverging behavior in the superfluid region. This difficulty can be removed if we allow for the presence of the superfluid phase in the trap by a mean field decoupling of the kinetic energy term and then perform the strong-coupling expansion about that state. We have carried out this procedure up to second order. The correction is everywhere small, and therefore controlled, but measurable, and the results are in good agreement with experiments.

Another important problem is to understand the universal scaling behavior of the Mott to superfluid quantum transitions at low temperatures. It has been suggested both theoretically \cite{TinHo1} and experimentally \cite{ChinCrit,Zhang2011,Hung2011} that the scaling properties can be verified by using real space density profiles at different temperatures. We have analyzed the scaling properties of our calculations in detail and find that they do not quite give universal scaling behavior for the vacuum to superfluid transition  in the regimes that have been studied experimentally, in that the non-scaling corrections are substantial, and the exactly scaling part is trivial. We show that infinite order resummation techniques need to be carried out and analyzed carefully to resolve these issues, as one might expect.

One way to approximately determine the temperature in experiments is by measuring its entropy \cite{Gerbier,Weiss,Blakie} before the optical lattice is applied. Generally, the entropy per particle is calculated for the homogeneous gas. Then the optical lattice is turned on adiabatically so that the total entropy as well as the entropy per particle remains unchanged. We have calculated the entropy up to second-order in $t/U$. From this calculation, we can generate plots of entropy per particle vs. temperature for different trap geometries so that it can serve as a look up table for estimating the temperature of the optical lattice system. There are other proposed ways for estimating the temperature of the optical lattice system as described in \cite{Zhou,Ketterle2,DeMarco}.

The rest of this paper is organized as follows. In Sec. II, we discuss the phase diagram of the Bose Hubbard model at different values of the parameters and temperature which will be useful to understand the coexisting phases which occur in experimental situations. In Sec. III, we discuss the strong-coupling perturbation technique about the mean-field result which proves to be a useful technique for calculating density profiles with coexisting phases. Next we compare our results with the finite temperature quantum Monte Carlo results as given in Ref. ~\onlinecite{Nandini}. In Sec. IV, we discuss the scaling properties of our calculations. In Sec. V, we discuss calculations of the entropy per particle which are useful for thermometry. Conclusions are presented in Sec. VI.


\section{The Bose Hubbard Model With A Harmonic Trap}

In the Bose-Hubbard model, bosonic atoms hop between nearest neighbour sites with energy $-t$, interact with an on site density-density repulsion $U$, feel a local energy determined by the local trap potential $V_{tr}|{\bf r}_j|^2$ and a global chemical potential $\mu$, with ${\bf r}_j$ the spatial coordinate of the $j$th lattice site. The Hamiltonian is ~\cite{Fisher}

\begin{eqnarray}
\nonumber
\mathcal{H}=&&-\sum_{jj'}t_{jj'}b^{\dagger}_{j}b_{j'}+\frac{U}{2}\sum_j n_{j}(n_{j}-1)-\mu \sum_j n_{j} \\
&&+V_{tr} \sum_j |\mathbf{r}_{j}|^2 n_j .
\label{eq1}
\end{eqnarray}
Here $b^{\dagger}_{j}$ and $b_{j}$ are bosonic creation and annihilation operators at site $j$ , respectively, which satisfy the commutation relations $[b_i,b_j^\dagger]=\delta_{ij}$ and all other commutators vanish. When $V_{tr} = 0$, at low enough temperatures, the model supports three phases (a) an incompressible Mott phase at $T=0$ when the filling is an integer; (b) a compressible normal phase for $T>0$ and integer filling, and $T>T_c(n)$ for non integer filling; and (c) a compressible superfluid phase for all noninteger fillings when $T<T_c(n)$. In the bulk, in one-dimension $T_c(n)=0$ and there is no superfuidity, and in two-dimensions $T_c(n)$ comes from the Kosterlitz-Thouless effect.

\begin{figure}[t]
\includegraphics[height=6.5cm,width=5.5cm]{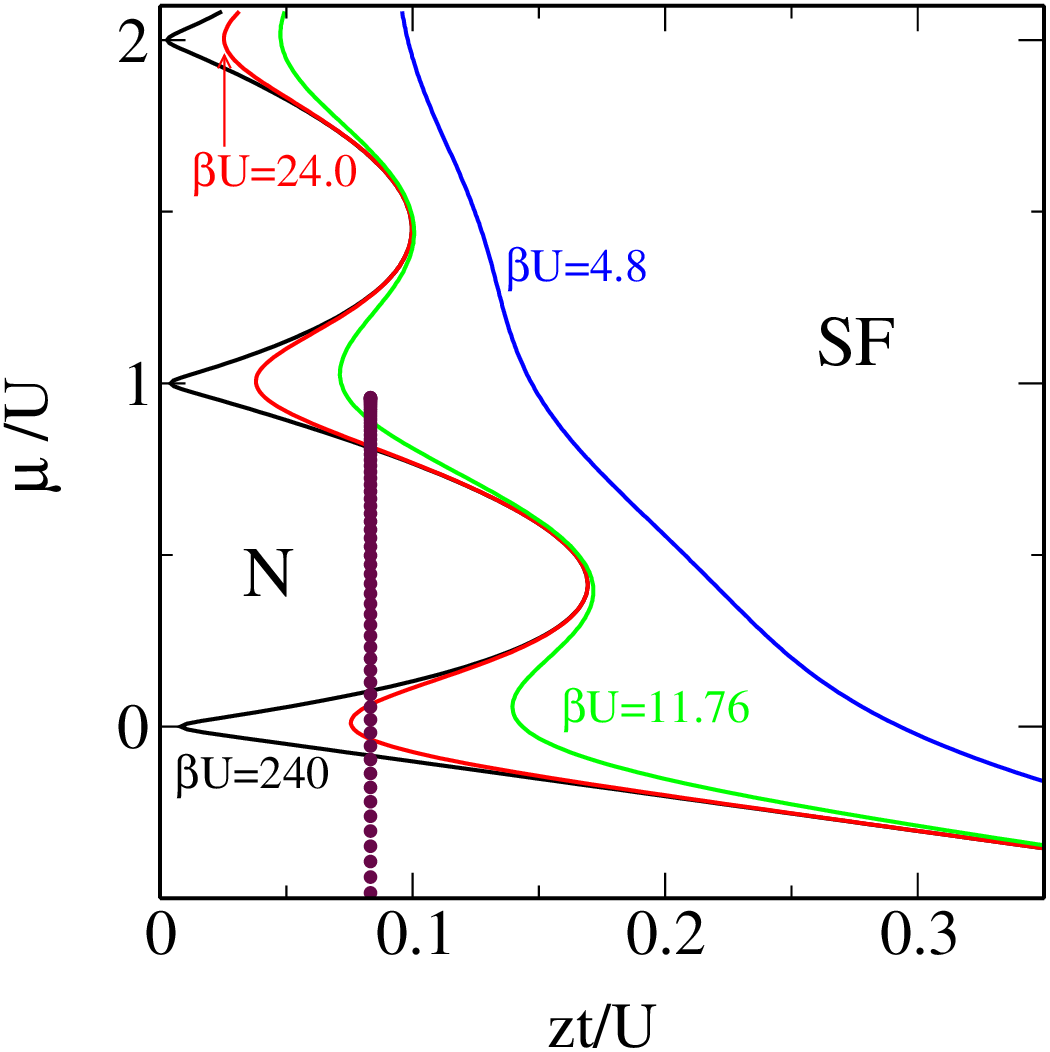}
\caption{(Color online) Phase diagram ($zt$ vs $\mu/U$) for the two-dimensional square lattice determined by the strong-coupling perturbation theory through second order in $(t/U)$ for different values of $\beta U$. The four different temperatures are $\beta U =240$ (black) $\beta U =24$ (red), $\beta U = 11.76$ (green), $\beta U = 4.8$ (blue). The dotted line shows the effective chemical potentials for a trap with trap size $r_d/a=7.45$ for a fixed value of $zt/U=0.0833$ radially outwards from the center (which would be used for a local density approximation calculation). The normal phase will become the Mott phase when $T=0$.}
\label{fig:graph1}
\end{figure}

The Mott phase transforms immediately to the normal phase as $T$ increases from zero (albeit with a very small compressibility when the temperature is much less than $U$). The phase diagram for a two-dimensional square lattice as obtained from simple mean-field theory \cite{Sheshadri} is shown in Fig.~\ref{fig:graph1}, and is discussed in more detail in  Sec. III. For a 2D square lattice, the coordination number, $z=4$. When $V_{tr} \ne 0$, [cf. Eq. (\ref{eq1})], in the local density approximation (LDA), one has a local chemical potential given by $\mu_j=\mu-V_{tr} |\mathbf{r}_{j}|^2$ which decreases radially outwards. The maroon dots in Fig. ~\ref{fig:graph1} show the effective chemical potentials $\mu_j$ at different lattice sites from the central site radially outwards for a fixed value of $zt/U$ in 2D. Fig. ~\ref{fig:graph2} shows the density and the order parameter as functions of the chemical potential. Hence, as one goes from the center to the periphery in the trap there results the coexistence of consecutive annular regions of Mott and superfluid phases leading to the well known ``wedding cake" density profiles \cite{Chin, Bloch2,Chin2} as shown in Fig. ~\ref{fig:graph3}.

The overall trap potential can be written as $V_{tr} |\mathbf{r}_{j}|^2=t|\mathbf{r}_{j}|^2 /|\mathbf{r}_{d}|^2$ which defines the characteristic length scale $r_d=\sqrt {t/V_{tr}}$, that governs the spatial variation of density, super fluid order parameter, {\it etc}. in the trap. Also this is a useful way to quantify the trap size. In our paper, we have written the trap size in terms of $r_d/a$, where $a$ is the lattice parameter.


\section{Strong-Coupling Expansion About The Mean Field Solution Including Superfluidity}

There have been many approaches to perform perturbation expansions in $t$. Here we build on the formalism of the strong-coupling expansion developed in Ref. \onlinecite{Freericks}. Our aim is to calculate the real space density profile of the bosons for benchmarking recent {\it in situ} measurements including the regions with superfluidity as accurately as possible. For this purpose, we allow for superfluidity by performing a mean field decoupling of the hopping term. Then we perform the strong-coupling perturbation expansion \cite{Freericks} about the mean-field solution. The hopping term can be re-written as:
\begin{equation}
\mathcal{H}_{hop}=-\sum_{jj'}t_{jj'}(\langle{b^{\dagger}_j\rangle}_{\tilde{\mathcal{H}}_{0j}}+\tilde{b}^{\dagger}_j)(\langle{b_{j'}\rangle}_{\tilde{\mathcal{H}}_{0j}}+\tilde{b}_{j'}) .
\label{eq9}
\end{equation}
Here the fluctuations from the average value of the creation and annihilation operators are defined as $\tilde{b}^{\dagger}_j\equiv[b^{\dagger}_j-\langle{b^{\dagger}_j \rangle}_{\tilde{\mathcal{H}}_{0j}}] $ and $ \tilde{b}_{j'}\equiv [b_{j'}-\langle{b_{j'} \rangle}_{\tilde{\mathcal{H}}_{0j}}]$, the local mean-field Hamiltonian $\tilde{\mathcal{H}}_{0j}$ being defined below in Eq. (\ref{eq11}).

The Hamiltonian in Eq. (\ref{eq1}) is then re-written as:

\begin{equation}
\mathcal{H}\equiv [\tilde{\mathcal{H}}_0+\sum_{jj'}\phi^*_jt^{-1}_{jj'}\phi_{j'}]+\tilde{\mathcal{H}}_I ,
\label{eq10}
\end{equation}

\begin{eqnarray}
\nonumber
\tilde{\mathcal{H}}_0 &&=\sum_j\left[ \frac{U}{2}n_j(n_j-1)-\mu_jn_j-\phi_jb^{\dagger}_j-\phi^*_jb_j\right] \\
&&\equiv\sum_j \tilde{\mathcal{H}}_{0j} .
\label{eq11}
\end{eqnarray}
Here $\phi_j\equiv \sum_{j'}t_{jj'}\langle{b_{j'} \rangle}_{\tilde{\mathcal{H}}_{0j}}$ is determined self-consistently from $\tilde{\mathcal{H}}_{0j}$. We can consider $\phi$ to be real for a simply connected trap, as in our case, without loss of generality.

\begin{eqnarray}
\nonumber
\tilde{\mathcal{H}}_I &&=-\sum_{jj'}t_{jj'}[b^{\dagger}_j-\langle{b^{\dagger}_j \rangle}_{\tilde{\mathcal{H}}_{0j}}][b_{j'}-\langle{b_{j'} \rangle}_{\tilde{\mathcal{H}}_{0j}}] \\
&&\equiv -\sum_{jj'}t_{jj'}\tilde{b}^{\dagger}_j\tilde{b}_{j'} .
\label{eq12}
\end{eqnarray}

In the strong-coupling limit ($t \ll U$), $\tilde{\mathcal{H}}_{0j}$ is diagonalized exactly and $\tilde{\mathcal{H}}_I$ is treated in perturbation theory.
All the average values are calculated using the basis of states that diagonalise $\tilde{\mathcal{H}}_{0j}$: $\tilde{\mathcal{H}}_{0j}|\tilde{n}\rangle_j=\tilde{\epsilon}_{j\tilde{n}}|\tilde{n}\rangle_j$. Finite temperature calculations are performed simply by taking thermal averages of the operators as $\langle{\textbf{A}_j\rangle}_{\tilde{\mathcal{H}}_0}=\sum_{\tilde{n}} {}_j\langle{\tilde{n}|\textbf{A}_j|\tilde{n}\rangle}_j\tilde{\rho}_{j\tilde{n}}$, where, $\tilde{\rho}_{j\tilde{n}}=e^{-\tilde{\epsilon}_{j\tilde{n}}}/\tilde{z}_j$ and $\tilde{z}_j=\textrm{Tr} e^{-\beta \tilde{\mathcal{H}}_{0j}}$ is the partition function at each site at zeroth order.


\subsection{The Perturbation Technique}

The perturbation technique is straightforward in terms of the basis of eigenvectors of $\tilde{\mathcal{H}}_0$. We expand the partition function as:

\begin{equation}
\mathcal{Z}=\textrm{Tr} e^{-\beta \mathcal{H}}=\tilde{\mathcal{Z}}_{0}[1+ \sum_{n} \tilde{\mathcal{Z}}^{(n)}] ,
\label{eq13}
\end{equation}
\begin{equation}
\tilde{\mathcal{Z}}^{(n)}=(-1)^{n} \int_0^{\beta} d\tau_n ... \int_0^{\tau_2} d\tau_1 \langle{\tilde{\mathcal{H}}_I(\tau_n)...\tilde{\mathcal{H}}_I(\tau_1)\rangle}_{\tilde{\mathcal{H}}_0} .
\label{eq14}
\end{equation}
Here $A(\tau)=e^{\tau\tilde{\mathcal{H}}_0}Ae^{-\tau\tilde{\mathcal{H}}_0}$ and $\tilde{\mathcal{Z}}_0=\prod_j \tilde{z}_j$. The first-order term is zero and the second-order term can be writen as

\begin{eqnarray}
\nonumber
\tilde{\mathcal{Z}}^{(2)} =&& \sum_{jj',j_1j'_1} t_{jj'} t_{j_1 j'_1} \int_0^{\beta}d\tau_2\int_0^{\tau_2} d\tau_1 \\
&&\langle{\tilde{b}_j^{\dagger}(\tau_2)\tilde{b}_{j'}(\tau_2)\tilde{b}_{j_1}^{\dagger}(\tau_1)\tilde{b}_{j'_1}(\tau_1)\rangle}_{\tilde{\mathcal{H}}_0} .
\label{eq15}
\end{eqnarray}
Nonvanishing terms arise either for $(a)$ $j'=j_1,j'_1=j$ or for $(b)$ $j'=j'_1,j_1=j$. The second combination is possible only in the presence of superfluidity, i.e. if $\phi_j \not=0$.

Finally, using the basis of eigenvectors of $\tilde{\mathcal{H}}_{0j}$, the second-order correction for the partition function can be written as:

\begin{eqnarray}
\nonumber
\tilde{\mathcal{Z}}^{(2)}&&= \sum_{jj'}|t_{jj'}|^2 \sum_{\tilde{n}\tilde{n}'}\sum_{\tilde{m}\tilde{m}'}\tilde{\rho}_{j\tilde{n}}\tilde{\rho}_{j'\tilde{n}'}\times \\
\nonumber
&& I_z^{(2)}(\tilde{\epsilon}_{j\tilde{m}}+\tilde{\epsilon}_{j'\tilde{m}'}-\tilde{\epsilon}_{j\tilde{n}}-\tilde{\epsilon}_{j'\tilde{n}'})\times \\
\nonumber
&&[|\langle{\tilde{m}|\tilde{b}_j|\tilde{n}\rangle}|^2|\langle{\tilde{m}'|\tilde{b}_{j'}^{\dagger}|\tilde{n}'\rangle}|^2 \\
&&+\langle{\tilde{n}|\tilde{b}_j^{\dagger}|\tilde{m}\rangle}\langle{\tilde{m}|\tilde{b}_j^{\dagger}|\tilde{n}\rangle}\langle{\tilde{n}'|\tilde{b}_{j'}|\tilde{m}'\rangle}\langle{\tilde{m}'|\tilde{b}_{j'}|\tilde{n}'\rangle}]
\label{eq17}
\end{eqnarray}
where, 
\begin{equation}
I_z^{(2)}(\epsilon)=\frac{\beta}{\epsilon}+\frac{e^{-\beta\epsilon}-1}{\epsilon^2} .
\label{eq18}
\end{equation}
For convenience of notation the subscripts $j$ on $|\tilde{n}\rangle$ and $|\tilde{m}\rangle$ and $j'$ on $|\tilde{n}'\rangle$  and $|\tilde{m}'\rangle$ have been suppressed.

The expectation value for any site-diagonal operator $\textbf{A}_j$ can be calculated (up to second-order in the perturbation) as :
\begin{eqnarray}
\nonumber
\langle{\textbf{A}_j\rangle}&&= \lbrace\langle{\textbf{A}_j\rangle}_{\tilde{\mathcal{H}}_0}-\int_0^{\beta} d\tau_1 \langle{\tilde{\mathcal{H}}_I(\tau_1) \textbf{A}_j\rangle}_{\tilde{\mathcal{H}}_0} \\
\nonumber
&&+ \int_0^{\beta} d\tau_2 \int_0^{\tau_2} d\tau_1 \langle{\tilde{\mathcal{H}}_I(\tau_2) \tilde{\mathcal{H}}_I(\tau_1) \textbf{A}_j\rangle}_{\tilde{\mathcal{H}}_0} \rbrace \times \\
&&\lbrace 1-\tilde{\mathcal{Z}}^{(2)} \rbrace .
\label{eq19}
\end{eqnarray}
The zeroth-order term is $\mathcal{A}^{(0)}_j\equiv \langle{\textbf{A}_j\rangle}_{\tilde{\mathcal{H}}_0}$. The first-order term again vanishes as $\langle{\tilde{b}^{\dagger}_{j_1}\rangle}$ or $\langle{\tilde{b}_{j_1}\rangle}$ will appear ``unpaired". The leading order term is the second-order term, given by:

\begin{eqnarray}
\nonumber
\mathcal{A}^{(2)}_j&&= \sum_{j_2 j'_2} \sum_{j_1 j'_1} t_{j_2 j'_2} t_{j_1 j'_1} \int_0^{\beta} d\tau_2 \int_0^{\tau_2} d\tau_1 \\
&&\langle{\tilde{b}^{\dagger}_{j_2}(\tau_2)\tilde{b}_{j'_2}(\tau_2) \tilde{b}^{\dagger}_{j_1}(\tau_1) \tilde{b}_{j'_1}(\tau_1) \textbf{A}_j\rangle}_{\mathcal{H}_0}-\mathcal{A}^{(0)}_j \tilde{\mathcal{Z}}^{(2)} .
\label{eq21}
\end{eqnarray}

Again in the diagonal basis of $\tilde{\mathcal{H}}_{0j}$, we can calculate the second-order term as:

\begin{eqnarray}
\nonumber
\mathcal{A}_j^{(2)} = &&\sum_{j'}\frac{|t_{jj'}|^2}{\tilde{z}_j\tilde{z}_{j'}} \sum_{\tilde{n},\tilde{n}^{'}} \sum_{\tilde{m},\tilde{m}^{'}} \sum_{\tilde{l}} \\
\nonumber
&&\lbrace  I^{(2)}(\beta ;\tilde{\epsilon}_{j\tilde{m}}+\tilde{\epsilon}_{j'\tilde{m}^{'}},\tilde{\epsilon}_{j\tilde{l}}+\tilde{\epsilon}_{j'\tilde{n}^{'}}, \tilde{\epsilon}_{j\tilde{n}}+ \tilde{\epsilon}_{j'\tilde{n}^{'}}) \\
\nonumber
&&[\langle{\tilde{n}|\tilde{b}^{\dagger}_j|\tilde{m}\rangle}\langle{\tilde{m}|\tilde{b}_j|\tilde{l}\rangle}\langle{\tilde{n}^{'}|\tilde{b}_{j'}|\tilde{m}^{'}\rangle}\langle{\tilde{m}^{'}|\tilde{b}^{\dagger}_{j'}|\tilde{n}^{'}\rangle} \\
\nonumber
&&+\langle{\tilde{n}|\tilde{b}_j|\tilde{m}\rangle}\langle{\tilde{m}|\tilde{b}_j|\tilde{l}\rangle}\langle{\tilde{n}^{'}|\tilde{b}^{\dagger}_{j'}|\tilde{m}^{'}\rangle}\langle{\tilde{m}^{'}|\tilde{b}^{\dagger}_{j'}|\tilde{n}^{'}\rangle} \\
\nonumber
&&+\langle{\tilde{n}|\tilde{b}^{\dagger}_j|\tilde{m}\rangle}\langle{\tilde{m}|\tilde{b}^{\dagger}_j|\tilde{l}\rangle}\langle{\tilde{n}^{'}|\tilde{b}_{j'}|\tilde{m}^{'}\rangle}\langle{\tilde{m}^{'}|\tilde{b}_{j'}|\tilde{n}^{'}\rangle} \\
\nonumber
&&+\langle{\tilde{n}|\tilde{b}_j|\tilde{m}\rangle}\langle{\tilde{m}|\tilde{b}^{\dagger}_j|\tilde{l}\rangle}\langle{\tilde{n}^{'}|\tilde{b}_{j'}|\tilde{m}^{'}\rangle}\langle{\tilde{m}^{'}|\tilde{b}_{j'}|\tilde{n}^{'}\rangle}] \\
&&(\langle{\tilde{l}|\textbf{A}_j|\tilde{n}\rangle}-\mathcal{A}^{(0)}_j\delta_{\tilde{l}\tilde{n}})\rbrace ,
\label{eq22}
\end{eqnarray}
\begin{eqnarray}
\nonumber
I^{(2)}(\beta;\epsilon_2,\epsilon_1,\epsilon_0)&&\equiv \frac{e^{-\beta\epsilon_0}}{(\epsilon_2-\epsilon_0)(\epsilon_1-\epsilon_0)} \\
\nonumber
&&+\frac{e^{-\beta\epsilon_1}}{(\epsilon_2-\epsilon_1)(\epsilon_0-\epsilon_1)} \\
&&+\frac{e^{-\beta\epsilon_2}}{(\epsilon_1-\epsilon_2)(\epsilon_0-\epsilon_2)} .
\label{eq23}
\end{eqnarray}
It can be shown that, although the separate terms in $I^{(2)}$ have singularities, they cancel so that $I^{(2)}$ never diverges.

One can calculate the real space distribution of atoms, which is an experimentally measurable quantity, by putting $\textbf{A}_j=\hat{N}_j$ in the expression given above, where $\hat{N}_j$ is the number operator at site $j$.

\begin{figure}[t]
\includegraphics[height=6cm,width=7cm]{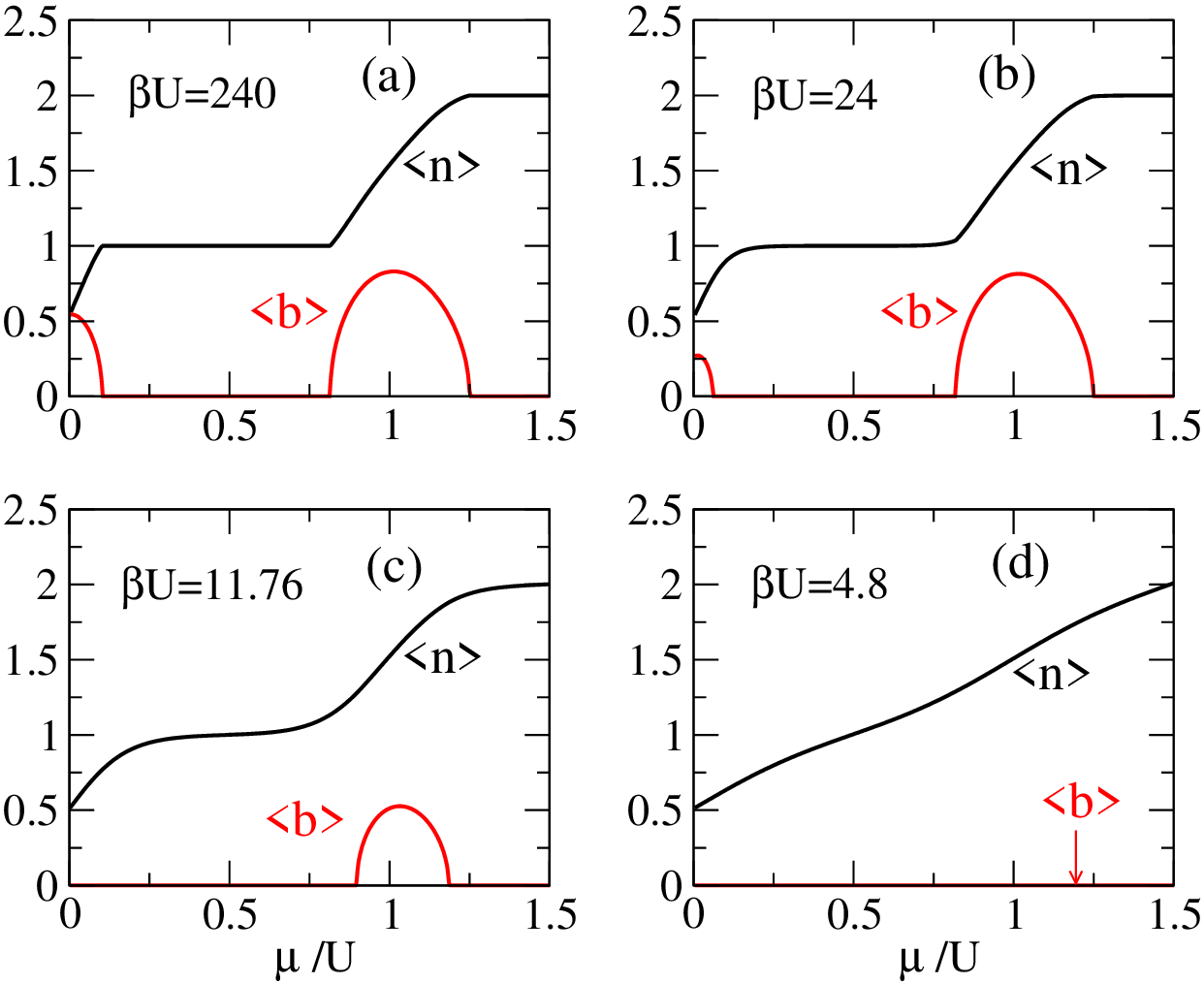}
\caption{(Color online) Number density $\langle{n\rangle}$ (black) and the order parameter $\langle{b\rangle}$ (red) for the homogeneous two-dimensional square lattice evaluated with the second-order strong-coupling expansion plotted as a function of the chemical potential for four different temperatures: $(a)$ $\beta U = 240$, $(b)$ $\beta U = 24$, $(c)$ $\beta U = 11.76$, $(d)$ $\beta U = 4.8$; $t/U=0.02084$.}
\label{fig:graph2}
\end{figure}


\subsection{Results and Analysis}

We have done the calculation in the homogeneous case as well as in the inhomogeneous case with a trap for a 2D lattice.  For the inhomogeneous lattice calculations, we have considered an overall harmonic trap of characteristic length $r_d/a = 7.45$. For the homogeneous case, we determine the order parameter $\phi$ self-consistently at a given value of chemical potential $\mu$ then calculate the number density and second-order correction to that. In our calculations, we have truncated the bosonic Fock space at each site while diagonalizing the Hamiltonian and kept a finite number of states (up to ten) in the occupation-number basis, which is reasonable because in our calculations we have explored parameter regimes that require occupancies only up to double occupancy. For the calculations in the presence of the trap, $\phi$ is first self-consistently determined at each site in the lattice using $\phi_j\equiv \sum_{j'}t_{jj'}\langle{b_{j'} \rangle}_{\tilde{\mathcal{H}}_{0j}}$, then the zeroth and second-order number densities are calculated in the lattice using Eqs. (\ref{eq22}) and (\ref{eq23}).

We use open boundaries for the lattice with a trap, with lattice sizes much bigger than the total number of bosons such that lattice sites near the boundaries are always empty. In this way, we can eliminate the effect of the boundary as long as the density at the edge of the boundary remains small enough that it can be neglected. The size of the lattice we have considered for our 2D calculation is $201\times201$ and the total number of bosons is about $8000$.

\begin{figure}[t]
\includegraphics[height=6cm,width=7cm]{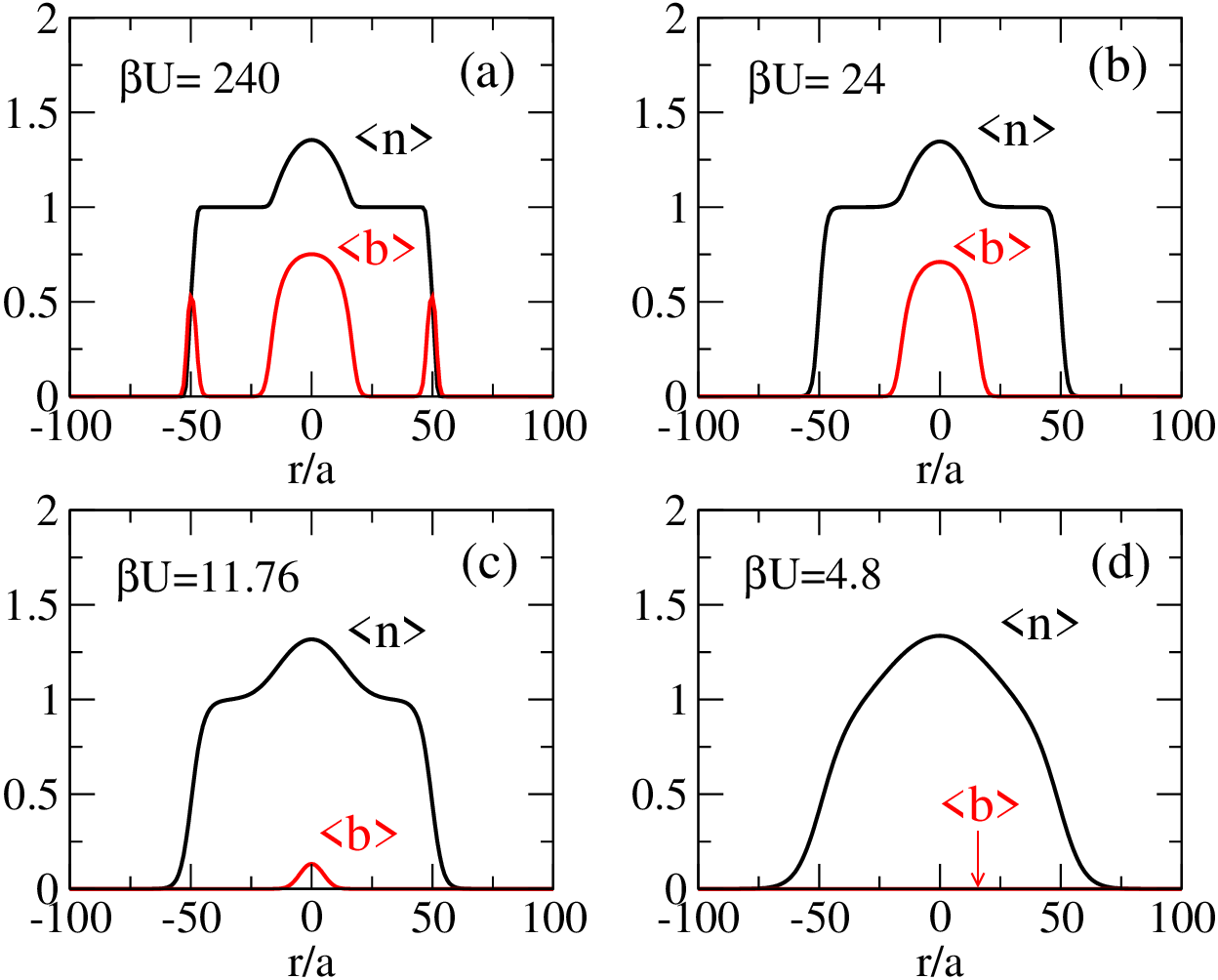}
\caption{(Color online) Density profile $\langle{n\rangle}$ (black) and the order parameter $\langle{b\rangle}$ (red) for the two-dimensional square lattice with a trap evaluated with the second-order strong-coupling expansion in the lattice with overall trap of size $r_d/a = 7.45$ for different temperatures: $(a)$ $\beta U = 240$, $(b)$ $\beta U = 24$, $(c)$ $\beta U = 11.76$, $(d)$ $\beta U = 4.8$; $t/U=0.02084$.}
\label{fig:graph3}
\end{figure}

 The results ($\langle{n\rangle}$ and $\langle{b\rangle}$) obtained from the calculations outlined above for the homogeneous case are shown in Fig.~\ref{fig:graph2} as a function of chemical potential, and with a trap are shown in Fig.~\ref{fig:graph3}. The second-order corrections to the results obtained from the mean-field decoupling of the original Hamiltonian are small for the range of parameters used in experiments. Hence, we may expect that this procedure has value in benchmarking the experiments.

\subsection{Comparison with LDA calculations}

\begin{figure}[t]
\includegraphics[height=7cm,width=8cm]{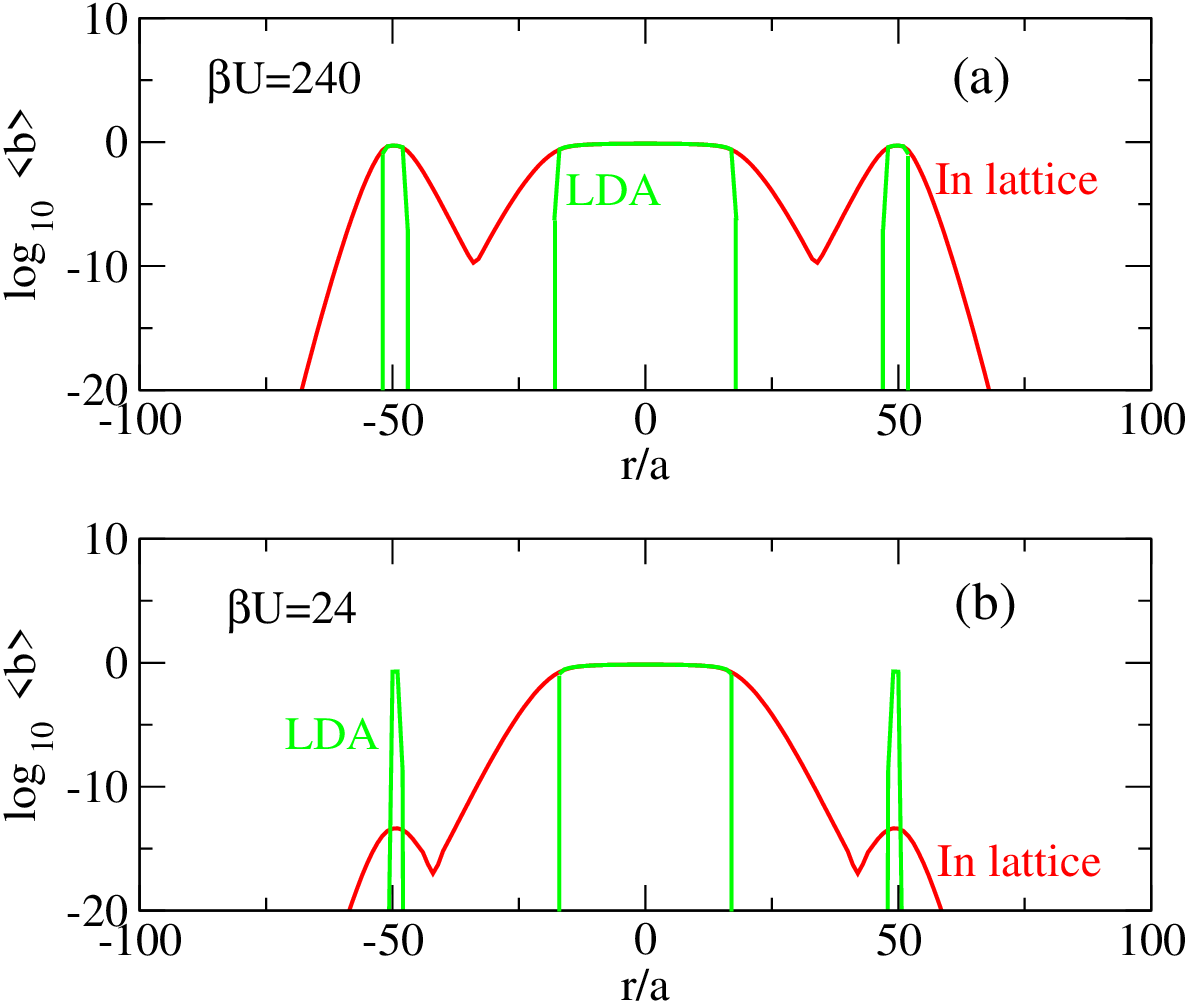}
\caption{(Color online) The logarithm of the order parameter ($\langle{b\rangle}$) along the trap for both LDA (green) and lattice (red) calculations at two different temperatures: $(a)$ $\beta U = 240$, $(b)$ $\beta U = 24$; $t/U=0.02084$; $r_d/a=7.45$.}
\label{fig:graph4}
\end{figure}

One can also calculate the density profile in the presence of the trap potential using the local density approximation (LDA) and the results for the homogeneous system. The results are generally in good agreement with the calculations done directly on the lattice for some parameter ranges ~\cite{Nandini}. However, there is one fundamental difference between the LDA results and the exact results in the context of the annular rings of coexistence of normal and superfluid phases. In the experimental situation and in the calculations done directly on the inhomogeneous lattice, if the superfluid region is present anywhere in the trap, the superfluid order parameter will in principle be finite everywhere inside the trap due to the proximity effect, though it is generally exponentially small in the regions where the LDA would predict the normal/Mott phase. Whereas in the LDA calculations, the superfluid order parameter is strictly zero in the normal/Mott regions. In Fig.~\ref{fig:graph4}, we have plotted the logarithm of the order parameter throughout the trap both for the LDA calculations (solid green lines) and calculations carried out directly on the 2D square lattice with a trap potential (solid red lines) at two different temperatures (a) $\beta U$ = 240 (Fig. ~\ref{fig:graph4}) and (b) $\beta U$ = 24 (Fig. ~\ref{fig:graph4}). [The data for $\langle{b\rangle}$ in Fig. ~\ref{fig:graph4}(a) and ~\ref{fig:graph4}(b) are the same as that of Fig. ~\ref{fig:graph3}(a) and ~\ref{fig:graph3}(b)]. These figures clearly show that there is no phase separation in the annular regions and superfluidity remains finite in the ``normal" region due to the proximity effect. They also show that there are regions where the LDA can greatly overestimate or underestimate the superfluid order parameter, so it is not wise to use it in the ordered phase for quantitative work.

\subsection{Comparison with Monte Carlo Results}

\begin{figure*}
\includegraphics[height=10cm,width=12cm]{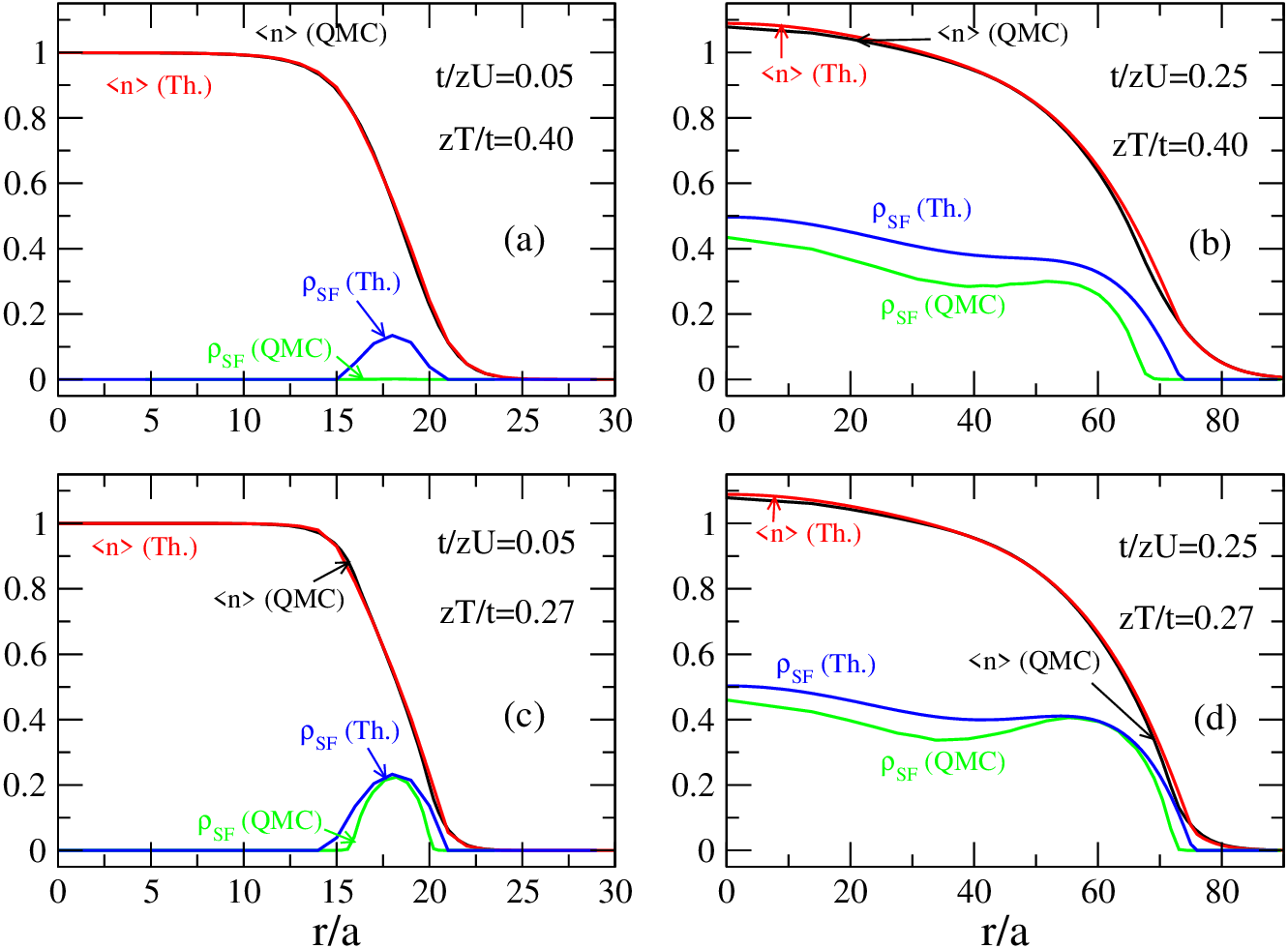}
\caption{(Color online) The density profile from QMC data (LDA) (black solid lines) \cite{Nandini} is compared with the strong-coupling calculation within LDA (red solid lines) for different temperatures and hopping. Superfluid density ($\rho_{SF}$) for QMC (solid green) and the strong-coupling calculation (solid blue) has also been plotted. For $(a)$ and $(c)$ the overall trap is given by $r_d/a=4.303$ and for $(b)$ and $(d)$ the same is given by $r_d/a=17.903$ as in Ref. \onlinecite{Nandini}.}
\label{fig:graph5}
\end{figure*}

We have compared our results with the quantum Monte Carlo data obtained from Zhou {\it et. al.} ~\cite{Nandini} using the LDA approximation. The comparison is shown in Fig. ~\ref{fig:graph5}. The calculation is done in 3D for four different cases: Fig. ~\ref{fig:graph5}(a) $t/zU=0.05$, $Tz/t=0.40$; ~\ref{fig:graph5}(b) $t/zU=0.25$, $Tz/t=0.40$; ~\ref{fig:graph5}(c) $t/zU=0.05$, $Tz/t=0.27$; ~\ref{fig:graph5}(d) $t/zU=0.25$, $Tz/t=0.27$, where the co-ordination number $z=6$ for the 3D calculation. The agreement is good for the density profile -- solid black lines show the QMC result and the solid red lines show the strong-coupling perturbation calculation. The comparison between the QMC result (solid green lines) and the strong-coupling perturbation calculation (solid blue) for the superfluid density, which we take to be $\rho_{SF}=\langle{b\rangle}^2$, is also shown in Fig. ~\ref{fig:graph5}. The quantitative agreement is good, although we find that the inclusion of the second-order contributions makes for better agreement than just mean-field theory, which tends to overestimate the superfluid order parameter as it neglects fluctuations. Higher order corrections not included in our calculation are primarily responsible for the discrepancy.

\section{Scaling Analysis for the Vacuum to Superfluid transition}

The scaled density $n(\mu,T)t/T$ (where temperature $T$ is expressed in energy units so that $\beta \equiv 1/T$) near the Mott-superfluid transition is expected to show universal scaling that is characteristic of the quantum critical point at low enough temperatures ~\cite{TinHo1,Hazzard}. The scaling theory requires the scaled density  to be a universal scaling function of the form

\begin{equation}
\frac{n(\mu,T)-n_{reg}(\mu,T)}{\frac{T}{t}}=F\left(\frac{\frac{\mu-\mu_c}{t}}{\frac{T}{t}}\right)=F\left(\frac{\mu-\mu_c}{T}\right) ,
\label{eq26}
\end{equation}
where $n_{reg}(\mu,T)$ is the non-singular or the regular part of the density (see below). The symbol $\mu_c$ is the critical chemical potential for the quantum (zero temperature) transition. Hence the standard prescription ~\cite{TinHo1} for verifying scaling is to plot the left hand side (LHS) of Eq. (\ref{eq26}) for different temperatures as a function of $\mu$ (or, in the spirit of the LDA, as a function of the radial distance in the trap), and the different curves should cross at $\mu_c$. Furthermore a plot of the LHS of Eq. (\ref{eq26}) as a function $(\mu -\mu_c)/T$ using that $\mu_c$ will then result in a collapse of the different curves to a universal curve. A key issue in this context is the choice of $n_{reg}$. In the prescription by Hazzard and Muller ~\cite{Hazzard}, $n_{reg}$ is the $t=0$ density, and in the prescription of Zhou and Ho ~\cite{TinHo1}, $n_{reg} = 0$ for the vacuum to superfluid transition.

 Within the spirit of the LDA, as mentioned above this feature can be, and has been, investigated experimentally by measuring the real space density in the system with the trap \cite{ChinCrit}. In this section we discuss the scaling properties of our calculations for the vacuum to superfluid transition and compare with the experimental results given in Refs. \onlinecite{ChinCrit} and \onlinecite{Zhang2011}. Our calculations are done for a 2D square lattice with an overall harmonic trap. The parameters are chosen to be the same as in the experiments \cite{ChinCrit} and are listed in the panels of the figures.


\subsection{Comparison with experimental scaled density}

\begin{figure*}
\includegraphics[height=9cm,width=14cm]{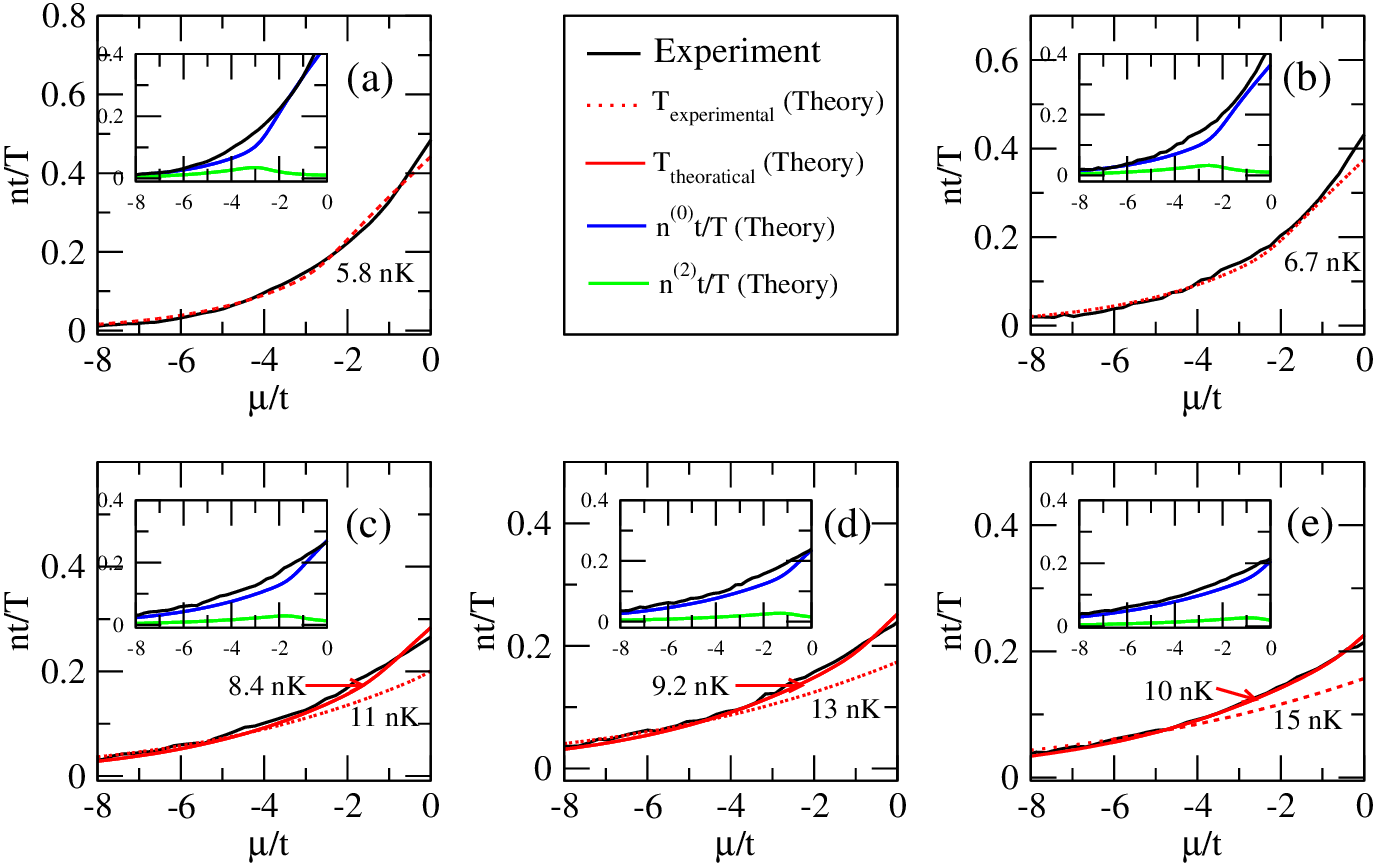}
\caption{(Color online) Comparison of the theoretical strong-coupling results with experimentally obtained scaled density profiles from Ref. \onlinecite{ChinCrit}. Solid black lines show the experimental data, red dotted lines show the theoretical calculations with the same temperatures as in the experimental analysis and red solid lines show the theoretical calculations at a different temperature that matches with the experimental data. The insets show the zeroth-order (solid blue) and the second-order (solid green) contributions of the scaled density separately. The overall harmonic trap (as given in Ref. \onlinecite{ChinCrit}) is $r_d/a=22.4$.}
\label{fig:graph6}
\end{figure*}

Fig.~\ref{fig:graph6} shows the comparison between our results and the experimentally measured scaled density \cite{ChinCrit} at different temperatures. Our calculations with second-order corrections included (red dots) match well with the experimental data (solid black lines) for temperatures 5.8 nK and 6.7 nK as shown in Fig. ~\ref{fig:graph6}(a) and ~\ref{fig:graph6}(b). For higher temperatures, the comparison is not as good if we use the experimentally quoted temperatures. For temperatures determined to be 11 nK(~\ref{fig:graph6}(c)), 13 nK (~\ref{fig:graph6}(d)) and 15 nK (~\ref{fig:graph6}(e)) in the experiments, theoretical calculations with temperatures lower than the temperatures estimated in the experiments match much better with the experimental data. As shown in Fig. ~\ref{fig:graph6} the ``11 nK" data match much better with theoretical calculations carried out at 8.2 nK, ``13 nK" data match with calculations at 9.4 nK and ``15 nK" match with calculations at 10 nK (solid red lines). The parameters used in the experiment are $t$=2.7 nK and $U$=15 nK, the total number of atoms are typically in the range of 4000 to 20000 and the plots are an average of different sets of data \cite{ChinCrit}. In the perturbative calculations, we have kept the chemical potential at the central site fixed. One reason for the difference could be that in the experiments the temperature of the system was determined by using a modified version of the mean-field treatment of a weakly interacting gas as in Ref. \onlinecite{TinHo2}, whereas our perturbative calculations correspond to an expansion in the strongly interacting regime ($U \gg t$). The strong-coupling calculation should work better at higher temperature, so this implies that the determination of the temperature at higher $T$ in the experiment may be problematic.

In the insets to Fig.~\ref{fig:graph6}, we have shown the zeroth-order (solid blue) and second-order (solid green) contributions to the scaled density along with the experimental data (solid black) which shows that the second-order contributions are significant for the experimental parameter regime and that their inclusion is essential for theory to match up with experiments.

\subsection{Scaling Properties of the Theoretical Results}

\begin{figure*}
\vspace{1cm}
\includegraphics[height=8.5cm,width=13cm]{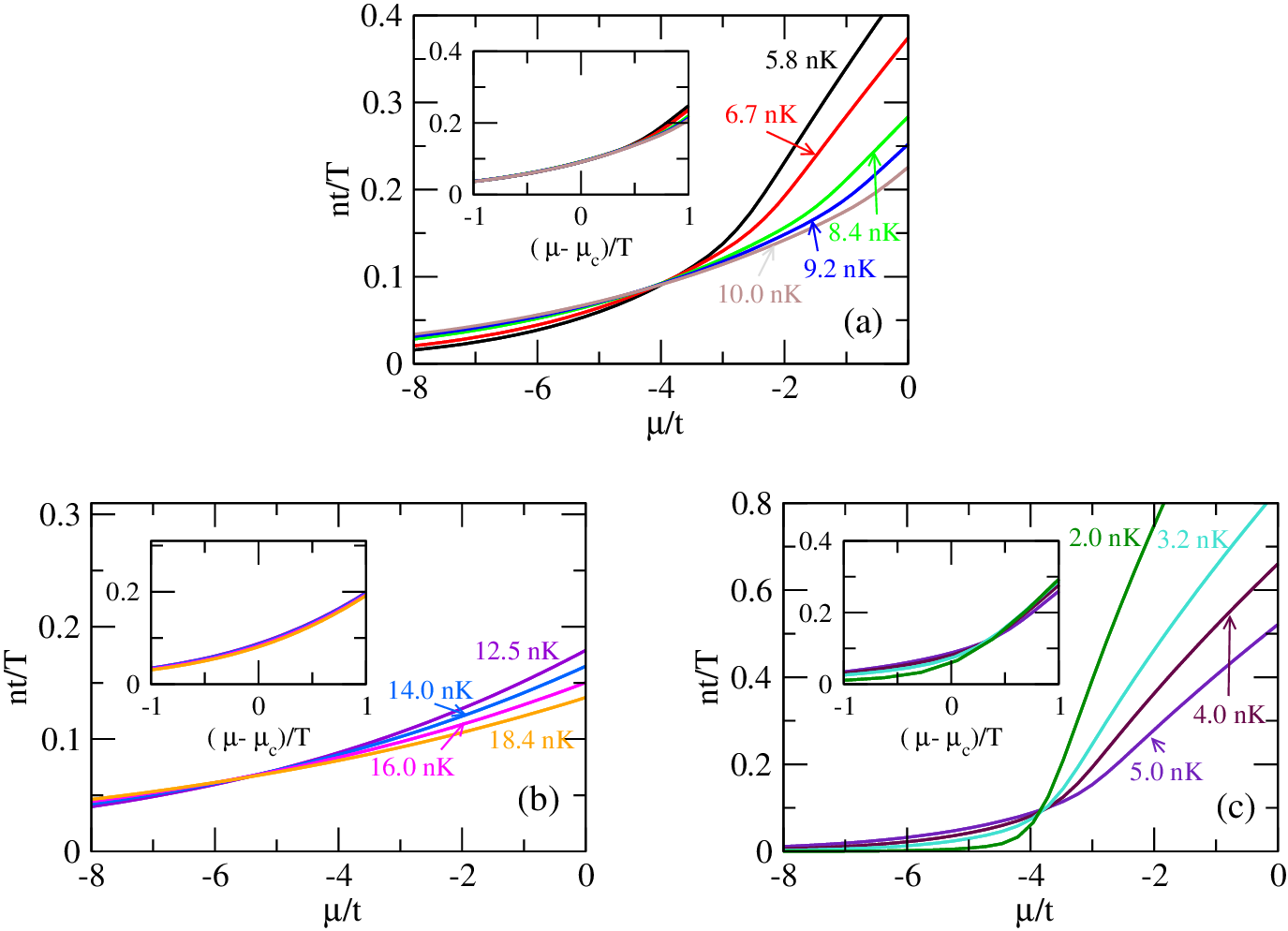} 
\caption{(Color online) Analysis of the scaling properties of the theoretical strong-coupling results. (a) Theoretical calculations for the scaled density for the experimental temperatures plotted versus the chemical potential showing the crossing at $\mu_c$. Inset: the scaled density as a function of $(\mu-\mu_c)/T$ illustrating the ``data collapse". Panels (b) and (c): similar to (a), but for a range of higher and lower temperatures respectively. The overall harmonic trap (as given in Ref. \onlinecite{ChinCrit}) is $r_d/a=22.4$.}
\label{new_all_theory}
\end{figure*}

Panel (a) of Fig.~\ref{new_all_theory} shows the calculated scaled densities for the different temperatures discussed above in one plot. The different curves do seem to intersect at $\mu_c/t \simeq 4.0$, and when re-plotted against $(\mu-\mu_c)/T$, do seem to show good data collapse, at least for $(\mu-\mu_c)/T < 0.4$, as seen in the inset of Fig.~\ref{new_all_theory}, panel (a). This would indeed seem to be a verification of universal quantum critical scaling, as suggested in the experimental analysis \cite{ChinCrit,Zhang2011}.

However, results from our calculations for higher (Fig.~\ref{new_all_theory}, panel (b)) and lower (Fig.~\ref{new_all_theory}, panel (c)) temperature ranges reveal some problems. For the higher temperature curves (Fig.~\ref{new_all_theory}, panel (b)) the common point of intersection seems to have shifted to the left of $\mu_c$, but the data collapse (inset of Fig.~\ref{new_all_theory}, panel (b)) while less accurate, seems to extend to a wider regime of $(\mu-\mu_c)/T$. But the lower temperature curves (Fig.~\ref{new_all_theory}, panel (c) and inset) show clear deviations from what is expected from scaling theory, which is bothersome.
 
Some insights into the problem with the scaling properties of our theoretical calculations (and what may be required to overcome them) can be gleaned by examining our calculations in a simplified context where only two bosonic states (occupancies 0 and 1) are retained at each site, in which case, as we show below, the strong coupling mean field density does not strictly have a nontrivial scaling form; in particular the value of the scaled density at the critical chemical potential ($\mu_c$) goes to zero exponentially with $\beta \mu_c$. Since our calculations are perturbative only up to second order about the mean-field results, they do not show universal scaling behavior either. We believe that the results discussed above, where up to 10 bosonic states were retained at each site, will have qualitatively similar features; hence their deviation from universal scaling at low temperatures.

The strong coupling mean field theory of the Bose Hubbard model in 2D for the vacuum to superfluid transition, with the simplification of retaining only bosonic states with occupation 0 and 1 at each site, can be carried out exactly analytically. In the homogeneous case (and hence within the LDA), near the vacuum to superfluid transition the average number density is then given by the following formula:
\begin{equation}
n(\mu,T) = \frac{\frac{2\phi^2}{\Delta (\Delta+|\mu|)}+\frac{\Delta+|\mu|}{2 \Delta}e^{-\beta \Delta}}{1+e^{-\beta \Delta}} ;
\label{eq24}
\end{equation}
where, $\Delta=\sqrt{\mu^2+4\phi^2}$ and $\phi(=-4t\langle{\hat{b}\rangle})$ are determined self-consistently using the following relation:
\begin{equation}
\phi = -\frac{4t\phi}{\Delta}\tanh(\frac{\beta \Delta}{2}) .
\label{eq25}
\end{equation}
It is straightforward to show from Eq.~(\ref{eq25}) that the chemical potential  for the finite temperature ``vacuum" to superfluid transition, $\mu_c (T)$, satisfies the equation:
\begin{equation}
\mu_c (T)=-4t+8te^{-\beta |\mu_c (T)|} .
\label{eq27}
\end{equation}
At low temperatures ($\beta$ large) this does not have any algebraic dependence on temperature. The quantum critical chemical potential is clearly  $\mu_c = - 4t$.

In the ``vacuum" phase, where the order parameter ($\phi$) vanishes, the number density is given by
\begin{equation}
n_{vac}(\mu,T)=\frac{e^{-\beta |\mu|}}{1+e^{-\beta |\mu|}} .
\label{eq28}
\end{equation}
Since $\phi$ is clearly zero at the critical chemical potential $\mu_c$ the value of $n_{vac}(\mu_c,T)$ or the scaled density $n_{vac}(\mu_c,T)t/T$ is different for different values of temperature. In other words, within the Zhou-Ho ~\cite{TinHo1} prescription and strong-coupling mean-field theory the LHS in Eq.~(\ref{eq26}) is not a function of $\frac{\mu-\mu_c}{T}$ alone. If one uses the Hazzard-Mueller ~\cite{Hazzard} prescription, the entirety of Eq.~(\ref{eq28}) has to be treated as $n_{reg}(\mu, T)$, whence the LHS of Eq.~(\ref{eq26}) is zero in strong-coupling mean-field theory.

In the superfluid phase, corresponding to $\phi \ne 0$, there is an additional contribution to $n(\mu, T)$ [cf. Eq. (\ref{eq24})] with an implicit temperature dependence coming from $\phi(T)$ and $\Delta(T)$ [through $\phi(T)$]. If we examine the self consistency equation [Eq. (\ref{eq25})] for small $\phi$ and large $\beta$, we get the following expression
\begin{equation}
\frac{8t\phi^2}{\mu^3}+\frac{8t}{\mu}e^{-\beta \mu}e^{-\frac{2\beta \phi^2}{\mu}} = 1+\frac{4t}{\mu} .
\label{eq29}
\end{equation}
The above equation shows that the temperature dependence of $\phi$ and hence of $\Delta$ within strong-coupling mean-field theory comes via an exponential dependence on $\beta \mu$ for low temperatures. For chemical potential values very close to the critical chemical potential $\mu_c$, the function $n(\mu,T)t/T$ therefore has $|\mu-\mu_c|/T$ dependence as well as $\mu_c/T$ dependence. It does not seem easy to define a function for the regular part of density $n_{reg}(\mu,T)$, which if we subtract from the density, $n(\mu,T)$ [in Eq. (\ref{eq24})] can remove the exponential dependence on $\mu_c/T$. The best one seems to be able to do is to work very close to $\mu_c$ and at low enough temperatures that the non-scaling components  are negligible, whence Eq. (\ref{eq26}) will be satisfied with the somewhat trivial scaling function given by $F[(\mu-\mu_c)/T]=0$ for the vacuum side of the transition and for the superfluid side the scaling function is given by $F[(\mu-\mu_c)/T]=(\mu-\mu_c)/8T$.

Even if we include the second-order correction to our scaled density, which is as given below for the ``vacuum" part, it does not formally show the scaling behavior in Eq.(~\ref{eq26}):
\begin{equation}
n^{(2)}_{vac}(\mu,T)=-\frac{t^2\beta^2}{2}\frac{e^{-\beta|\mu|}}{(1+e^{-\beta|\mu|})^2}\left[1+\frac{2e^{-\beta|\mu|}}{1+e^{-\beta|\mu|}}\right] .
\label{eq29I}
\end{equation}

\begin{figure*}[t]
\includegraphics[height=9cm,width=14cm]{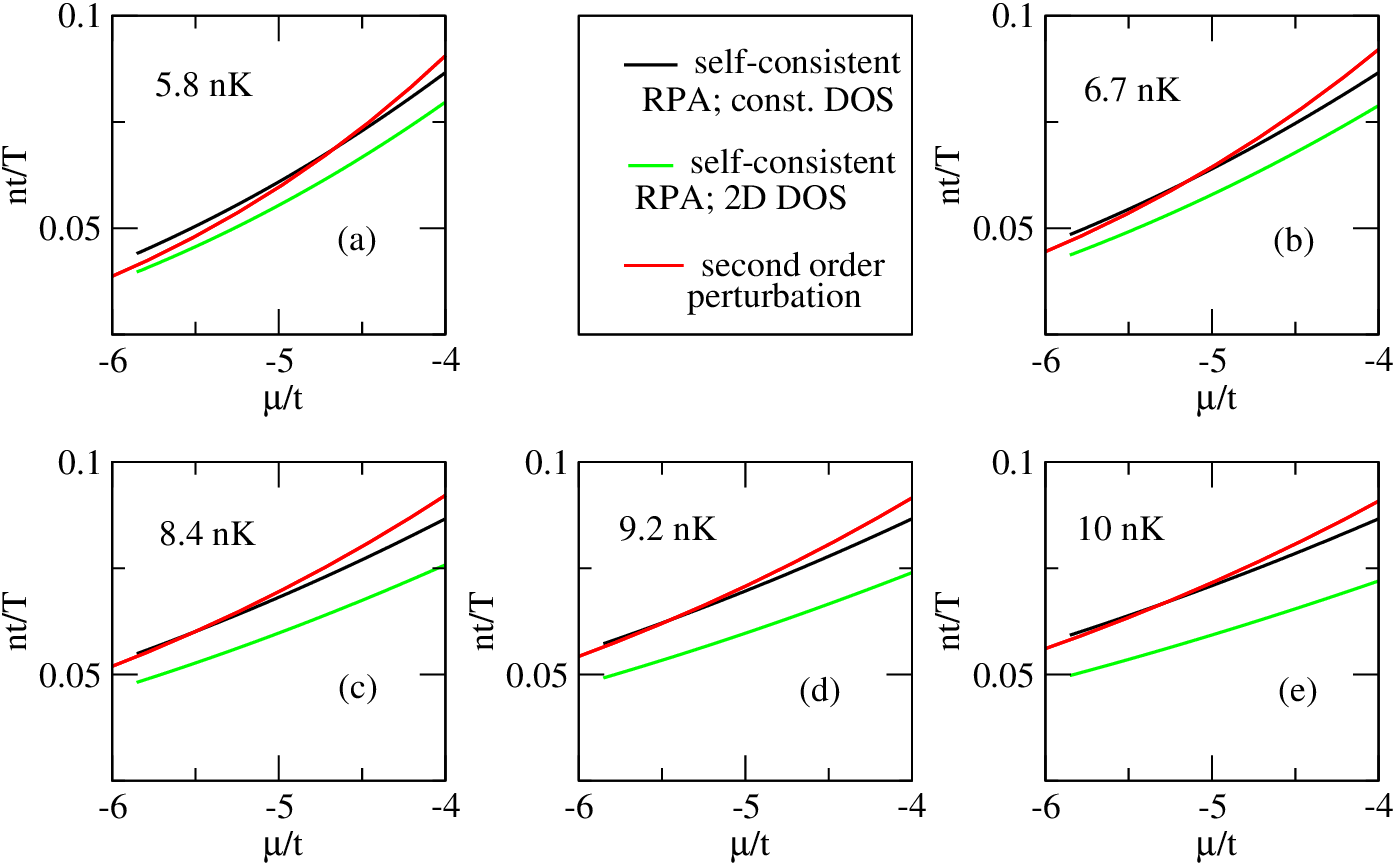}
\caption{(Color online) Comparison between the scaled density calculated from self-consistent RPA with a constant DOS (solid black), with a tight-binding DOS for 2D square lattice (solid blue) and second-order strong-coupling perturbation theory (solid red) on the ``vacuum" side for the same temperatures as those that lead to the best fit of the perturbative calculations to the experimental data. }
\label{fig:graph6I}
\end{figure*}

In view of the good agreement between our calculations, which included the non-scaling terms, and the experiments, and the fact that numerically they seem to show scaling at least for an intermediate range of temperatures, it is worth investigating whether these problems can be cured by performing calculations that involve infinite order summations of subsets of perturbative fluctuation corrections to strong coupling mean field theory. The simplest such summation is the random phase approximation (RPA). A {\it self-consistent} RPA calculation of the density in the ``vacuum" state, where we approximate the single boson tight-binding density of states (DOS) as a constant within the band of width $8t$ ($g(\epsilon)=1/8t$), can be carried out exactly [for the simplified case where only two bosonic states (occupancies 0 and 1) are retained at each site], and yields the following expression which is indeed of the scaling form in Eq. (\ref{eq26}).
\begin{equation}
\frac{n_{RPA}^{sc}(\mu,T)t}{T}=\frac{1}{8}\ln\lbrack{1+e^{-\beta(\mu_c-\mu)}\rbrack} .
\label{eq29II}
\end{equation}
But, if we carry out the self-consistent RPA for the ``vacuum" state using the density of states with energy dependence, such as that of the tight-binding band on a square lattice, there are corrections which do not have the universal scaling form.

For a general case where the DOS is given by $g(\epsilon)=\sum_p \frac{g_p}{p!} (\epsilon+4t)^p$, one finds that the self-consistent RPA density $n$ satisfies the following equation:
\begin{equation}
\frac{nt}{T}=\sum_p \frac{1}{\alpha_p}\left[\frac{T/t}{1-2n}\right]^p \mathcal{B}_{p+1}(x) ,
\end{equation}
where
\begin{equation}
x \equiv e^{\frac{(\mu-\mu_c)}{T}}e^{-\frac{8nt}{T}} .
\end{equation}
Here we have written $g_p=1/(\alpha_p t^{p+1})$, where $\alpha_p$ are numerical coefficients, exponentially small terms of order $\exp(-8t/T)$ or smaller have been neglected, and $\mathcal{B}_{p+1}(x)$ are the Bose-Einstein functions defined as:
\begin{equation}
\mathcal{B}_{p+1}(x) = \int_0^{\infty} \frac{y^p dy}{x^{-1}e^y-1} .
\end{equation}
While an approximation of keeping just the first term in this series leads to a density which is indeed of the scaling form, the other terms spoil it, because of the appearance of the $T/[t(1-2n)]$ factors.

In Fig.~\ref{fig:graph6I}, we have compared the different self-consistent RPA calculations with the scaled density calculated up to second-order in the perturbation series. The RPA scaled density calculated from the constant density of states, which shows universal scaling, is surprisingly close to the scaled density from strong-coupling perturbative calculation. On the other hand, for the actual 2D density of states, the agreement with the perturbative calculation is not as good. In the figure shown (Fig.~\ref{fig:graph6I}), we can see that among the five different temperatures as used in the experiments, the lowest one ( 5.8 nK) has the least discrepancy between the scaled density calculated including second-order perturbation corrections and the self-consistent RPA calculation with the actual DOS. Note also that the curvatures of the density versus chemical potential curves as obtained from the self-consistent RPA calculations are noticeably different from those from the calculations that include the second-order perturbation corrections. This shows that the terms beyond the RPA, eg., those that include second order self energy corrections, might be important in this parameter regime for properly sorting out issues connected with universal quantum critical scaling. Clearly a more careful analysis of both theory and experiment would seem to be needed to understand the extent to which universal scaling is obeyed.


\section{Entropy per particle}

In cold atoms systems, the average entropy per particle can be employed as a thermometer since the entropy is a state variable thermodynamically conjugate to the temperature and can be used as a temperature scale. Nevertheless, it is often useful to try to estimate the temperature of these systems in order to better understand their behavior. In our formalism, we can straightforwardly calculate the entropy of the homogeneous and lattice system as a function of  temperature up to second order in $t/U$ using the relation $S=(\langle{\mathcal{H}\rangle}-F)/T$, where $F$ is the free energy of the system and $\langle{\mathcal{H}\rangle}$ is the average value of energy at a particular temperature $T$. The free energy is easily calculated from the partition function, $F=- T \ln \mathcal{Z}$. Up to second-order in the perturbation series, the free energy can be written as,

\begin{eqnarray}
F&&=- T \sum_j \ln \tilde{z}_j- T \mathcal{Z}^{(2)}+\sum_{jj'}\phi^*_jt^{-1}_{jj'}\phi_{j'} \\
\nonumber
&&= \sum_j \mathcal{F}^{(0)}_j+\mathcal{F}^{(2)}+\sum_{jj'}\phi^*_jt^{-1}_{jj'}\phi_{j'} .
\label{eq30}
\end{eqnarray}

To calculate the average value of the energy up to second-order in the perturbation series, we go back to the total Hamiltonian given in Eq. (\ref{eq10}). We calculate the average value order by order in the perturbation series for both the diagonal and the perturbative parts of the total Hamiltonian.

The zeroth-order entropy is calculated from the zeroth-order contribution of the diagonal part of the Hamiltonian [in Eq. (\ref{eq11})] and the zeroth-order free energy ($\sum_j \mathcal{F}^{(0)}_j$). The constant in both the average energy and free energy expressions (given by $\sum_{jj'}\phi^*_jt^{-1}_{jj'}\phi_{j'}$) cancels out as we take the difference between the two.

\begin{equation}
S^{(0)}=\sum_j s^{(0)}_j = \sum_j (\langle{\tilde{\mathcal{H}}_{0j}\rangle}^{(0)}-\mathcal{F}_j^{(0)})/T .
\label{eq32}
\end{equation}

The second-order correction to the average value of energy also has two parts. One is from the second-order correction of $\langle{\tilde{\mathcal{H}}_{0j}\rangle}$ given by $\langle{\tilde{\mathcal{H}}_{0j}\rangle}^{(2)}$ and another from the first-order correction to $\langle{\tilde{\mathcal{H}}_{I}\rangle}$ which is also second-order in $t$, and is given by $\langle{\tilde{\mathcal{H}}_I\rangle}^{(1)}=\int_0^{\beta} d\tau_1 \langle{\tilde{\mathcal{H}}_{I}(\tau_1)\tilde{\mathcal{H}}_{I}\rangle}_{\tilde{\mathcal{H}}_0}  $. Combining the second-order free energy and the second-order contributions to the average energy the second-order correction to the entropy is given by:

\begin{equation}
S^{(2)}=(\sum_j \langle{\tilde{\mathcal{H}}_{0j}\rangle}^{(2)}+\langle{\tilde{\mathcal{H}}_I\rangle}^{(1)}-\mathcal{F}^{(2)})/T .
\label{eq33}
\end{equation}

The second-order correction to $\langle{\tilde{\mathcal{H}}_{0j}\rangle}$ can be calculated using the expressions in Eq. (\ref{eq22}), in the basis of eigenvectors of $\tilde{\mathcal{H}}_{0j}$. The first-order correction to $\langle{\tilde{\mathcal{H}}_{I}\rangle}$ can be calculated following the same procedures as in the calculation of $\mathcal{Z}^{(2)}$ and, is given by :

\begin{eqnarray}
\nonumber
\langle{\tilde{\mathcal{H}}_I\rangle}^{(1)}&&= \sum_{jj'}|t_{jj'}|^2 \sum_{\tilde{n}\tilde{n}'}\sum_{\tilde{m}\tilde{m}'}\tilde{\rho}_{j\tilde{n}}\tilde{\rho}_{j'\tilde{n}'}\times \\
\nonumber
&& I^{(1)}(\tilde{\epsilon}_{j\tilde{m}}+\tilde{\epsilon}_{j'\tilde{m}'}-\tilde{\epsilon}_{j\tilde{n}}-\tilde{\epsilon}_{j'\tilde{n}'})\times \\
\nonumber
&&[|\langle{\tilde{m}|\tilde{b}_j|\tilde{n}\rangle}|^2|\langle{\tilde{m}'|\tilde{b}_{j'}^{\dagger}|\tilde{n}'\rangle}|^2 \\
&&+\langle{\tilde{n}|\tilde{b}_j^{\dagger}|\tilde{m}\rangle}\langle{\tilde{m}|\tilde{b}_j^{\dagger}|\tilde{n}\rangle}\langle{\tilde{n}'|\tilde{b}_{j'}|\tilde{m}'\rangle}\langle{\tilde{m}'|\tilde{b}_{j'}|\tilde{n}'\rangle}]
\label{eq34}
\end{eqnarray}
where $I^{(1)}$ is given by :

\begin{equation}
I^{(1)}(\epsilon)=-\frac{e^{-\beta\epsilon}-1}{\epsilon} .
\label{eq35}
\end{equation}

If we collect all these terms in the second-order correction to the entropy, we notice that only $\langle{\tilde{\mathcal{H}}_{0j}\rangle}^{(2)}$ is defined on one site; the other two terms involve pairs of sites. Hence the second-order correction to entropy can be written in the form $S_2=\sum_{jj'} s^{(2)}_{jj'}$, i.e, a sum of contributions from each nearest neighbour  pair of sites. At very low temperatures, it is straightforward to see that the dominant part of the second-order entropy correction comes from the region in the lattice which has the chemical potential close to zero.


\subsection{Temperature Calibration}

\begin{figure}[t]
\includegraphics[height=6cm,width=8cm]{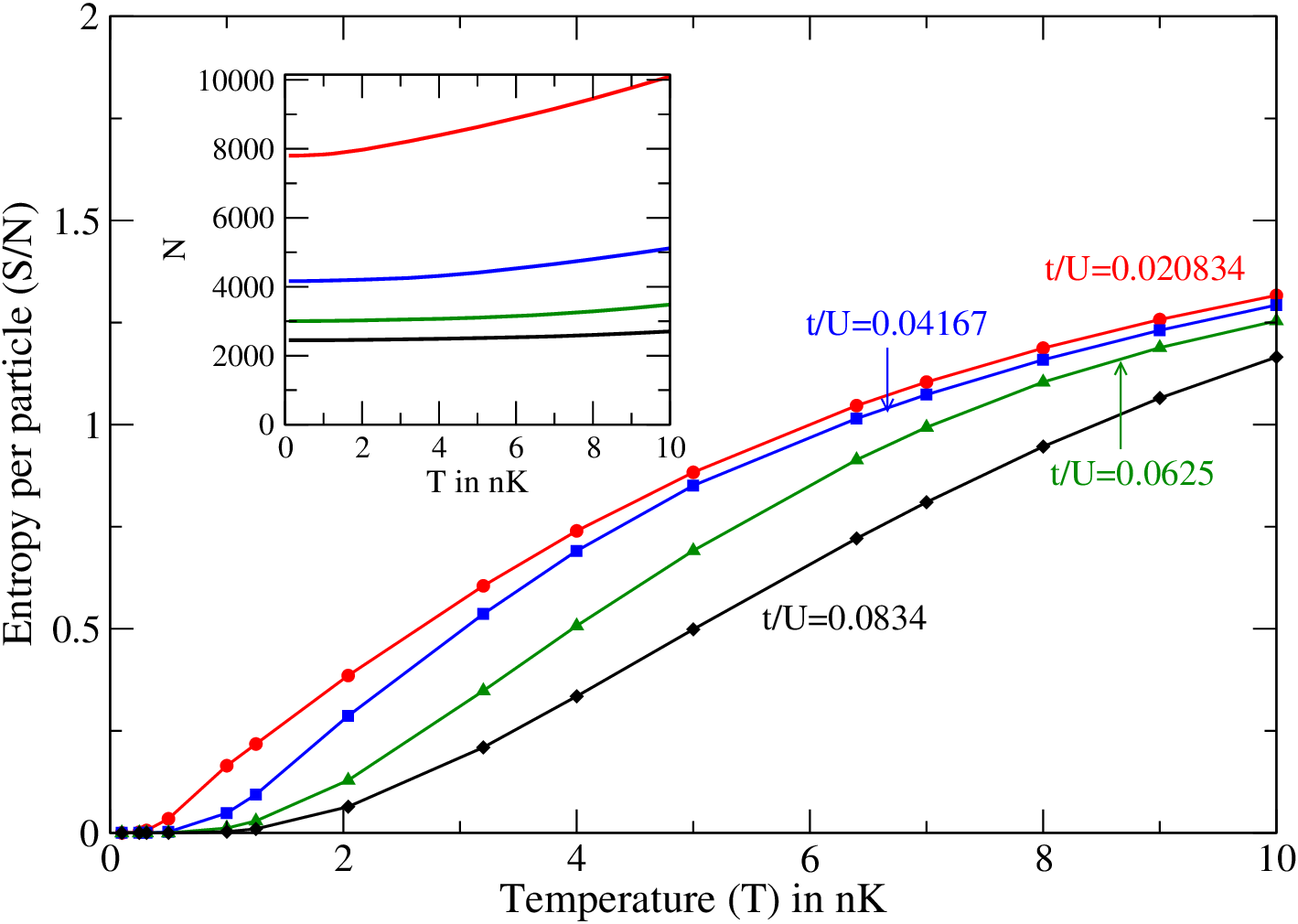}
\caption{(Color online) Entropy per particle plotted against the temperature for four different $t/U$ values. This calculations are done for a 2D square lattice with the chemical potential at the center fixed, and the overall trap potential is $r_d/a=7.45$. The inset shows how the total number of particles vary with the temperature for a given $t/U$. For the inset, the color codes are the same as the main plot.}
\label{fig:graph7}
\end{figure}

As we have stated earlier, in optical lattice experiments one commonly uses entropy per particle as the temperature scale. Our calculations of the entropy per particle presented above helps us to generate temperature calibration curves which can be used to estimate the temperature of the system during experiments, as has already been done for fermions \cite{Esslinger}. In Fig.~\ref{fig:graph7}, we have shown the calculated entropy per particle including the second-order correction as a function of temperature for different $t/U$ values for a 2D system. For different harmonic traps ($V_{tr}/U$), the entropy per particle curves with same chemical potential at the center and for the same hopping ($t/U$) essentially coincide with each other. This can be explained within LDA by following the same analysis \footnote{We write the total entropy ($S$) and total number of particles ($N$) as a sum over the quantity at each site as $S=\sum_i s_{2D}(\mu_0-V_Tr_i^2,t/U,T)$ and $N=\sum_i n_{2D}(\mu_0-V_Tr_i^2,t/U,T)$. In the continuum limit we can write, $N=\frac{\pi}{V_T}\int_{-\infty}^{\mu_0} d\mu_r n_{2D} (\mu_r,t/U,T) \equiv \frac{\pi}{V_T}I(\mu_0,t/U,T)$ and $S=\frac{\pi}{V_T}\int_{-\infty}^{\mu_0} d\mu_r s_{2D} (\mu_r,t/U,T) \equiv \frac{\pi}{V_T}J(\mu_0,t/U,T)$. The $V_T$ dependence is the same in both the cases and cancels out when we take the ratio.} as given in Ref. ~\onlinecite{Batrouni}.

\section{Conclusion}

In this paper, we have showed how a strong-coupling expansion about the mean-field can be developed for the Bose Hubbard model and gives useful results to benchmark the experiments being carried out in cold atom systems. The density profiles in the presence of a trap from our calculations are in good agreement with QMC results. Also, the scaled density curves from the experiments are in good agreement with our calculation including the second-order corrections. We have presented a detailed scaling analysis of our calculations for the vacuum to superfluid transition which suggests that an experimental verification of the expected universal scaling properties of the appropriate quantum critical point requires a more careful analysis of both theory and experiment. We hope to address this issue in future work. Finally, we have presented calculations for the entropy per particle vs. temperature which will be useful for the estimation of temperature in the experiments.


\begin{acknowledgments}
This work was supported by a MURI Grant from the Air Force Office of Scientific Research No. FA9559-09-1-0617. The Indo-US collaboration was supported by the Indo-US Science and Technology Forum under Joint Center Grant No. JC-18-2009 (ultracold atoms). J.K.F. also acknowledges the McDevitt bequest at Georgetown, and H.R.K. also acknowledges the support of the Department of Science and Technology of India. We thank Dr. Nandini Trivedi and Qi Zhou for providing us with their QMC data. We also thank Dr. Cheng Chin and Xibo Zhang for sharing their experimental data with us, for scaling analysis. 
\end{acknowledgments}


\bibliography{coldbosonpaper.bib}

\end{document}